\documentclass[fleqn,10pt]{SelfArx} 

\usepackage[english]{babel} 

\usepackage{lipsum} 

\definecolor{color1}{RGB}{0,0,90} 
\definecolor{color2}{RGB}{0,20,20} 

\usepackage{hyperref} 
\hypersetup{hidelinks,colorlinks,breaklinks=true,urlcolor=color2,citecolor=color1,linkcolor=color1,bookmarksopen=false,pdftitle={Title},pdfauthor={Author}}

\PaperTitle{Application of Stochastic and Deterministic Techniques for Uncertainty Quantification and Sensitivity Analysis of Energy Systems} 

\Authors{Majdi I. Radaideh\textsuperscript{1}*, Mohammad I. Radaideh\textsuperscript{2}} 
\affiliation{\textsuperscript{1}\textit{Department of Nuclear, Plasma, and Radiological Engineering, University of Illinois at Urbana Champaign, Urbana, Illinois 61801, United States}} 
\affiliation{\textsuperscript{2}\textit{Department of Mechanical Engineering, Jordan University of Science and Technology, P.O.Box 3030, Irbid 22110, Jordan}} 
\affiliation{*\textbf{Corresponding author}: radaide2@illinois.edu} 

\Keywords{Uncertainty Quantification, Sensitivity Analysis, Variance Decomposition, Fuel Cell, Stochastic Methods} 
\newcommand{\keywordname}{Keywords} 

\Abstract{Sensitivity analysis (SA) and uncertainty quantification (UQ) are used to assess and improve engineering models. In this study, various methods of SA and UQ are described and applied in theoretical and practical examples for use in energy system analysis. This paper includes local SA (one-at-a-time linear perturbation), global SA (Morris screening), variance decomposition (Sobol indices), and regression-based SA. For UQ, stochastic methods (Monte Carlo sampling) and deterministic methods (using SA profiles) are used. Simple test problems are included to demonstrate the described methods where input parameter interactions, linear correlation, model nonlinearity, local sensitivity, output uncertainty, and variance contribution are explored. Practical applications of analyzing the efficiency and power output uncertainty of a molten carbonate fuel cell (MCFC) are conducted. Using different methods, the uncertainty in the MCFC responses is about 10\%. Both SA and UQ methods agree on the importance ranking of the fuel cell operating temperature and cathode activation energy as the most influential parameters. Both parameters contribute to more than 90\% of the maximum power and efficiency variance. The methods applied in this paper can be used to achieve a comprehensive mathematical understanding of a particular energy model, which can lead to better performance.}

\begin{document}

\flushbottom 
\maketitle 
\tableofcontents 
\thispagestyle{empty} 


\section{Introduction}

Uncertainty Quantification (UQ) and Sensitivity Analysis (SA) are two essential components in scientific computing and engineering design \cite{smith2013uncertainty,le2010spectral}. UQ is a broad area of research concerned with characterization and reduction of uncertainties in both computational and experimental models. Uncertainty propagation (i.e. forward UQ) of system inputs to quantify their effect on system response is a typical UQ problem. Uncertainty propagation helps to characterize system performance when the system input parameters are unknown or have uncertainty inherent in their measurement. The uncertainty in input parameters is propagated through the scientific model (i.e. analytical, computer, semi-empirical, etc.) and the uncertainty in the response is quantified. This uncertainty in the response can be used to asses the confidence that should be placed on the result. Two main categories of methods are used to perform forward UQ (or simply UQ for convenience in this study). The first category is deterministic-based \cite{smith2013uncertainty, putko2002approach} which relies on determining first or second order sensitivity profiles of the input parameters, and then combining these profiles with the variance-covariance matrix of the input parameters. The second category is stochastic-based (or Monte Carlo) \cite{smith2013uncertainty,helton2006survey}, which relies on probability theory where the uncertainty of the input parameters is characterized by a random probability density function (e.g. normal, beta, lognormal, etc.). Random samples are generated from the parameter distributions and these samples are propagated through the scientific model. Samples can be generated randomly or using efficient Latin hypercube sampling techniques \cite{helton2003latin}. The first and second statistical moments (e.g. mean, variance) of the output distribution can be determined, which quantify response uncertainty. A wide range of applications of UQ methods can be found in literature. For example, in the context of computational fluid dynamics \cite{putko2002approach}, radioactive waste disposal \cite{helton1993uncertainty}, surrogate models \cite{le2010spectral}, environmental and biological systems \cite{isukapalli1998stochastic}, and many more. 

SA maps the relationship between model response and its input parameters. SA helps the analyst to identify and rank the input parameters with the largest influence on model response. A plethora of sensitivity analysis methods is available in literature. A comprehensive review of the recent SA methods can be found in \cite{borgonovo2016sensitivity, christopher2002identification}. SA is directly connected to UQ because deterministic-based UQ is performed through the sensitivity profile as obtained from SA. In addition, input parameters that both have large uncertainty and sensitivity are expected to have a large overall impact on the system's performance. SA can be classified in different forms, but four main categories will be highlighted due to their relevancy to this study: local methods \cite{saltelli2000sensitivity}, regression-based methods \cite{helton2006survey}, screening methods \cite{morris1991factorial}, and variance-based or global methods \cite{saltelli2008global}. In local methods, a single input parameter is perturbed with a small perturbation factor, while the other input parameters remain unchanged, and the effect of this perturbation on the output is determined. Finite difference \cite{lenhart2002comparison} and adjoint-based methods \cite{kiedrowski2013adjoint} are the most common local methods. Screening methods aim to identify the influential parameters in the model by performing a thorough exploration of the input space as compared to regular local methods. These methods can identify any nonlinearity in the model, while maintaining a minimal number of model evaluations. Morris screening \cite{morris1991factorial} constructs an $n_x$-dimensional grid ($n_x$ is number of input parameters), which covers the input range, and the parameter sensitivity is explored by repeating analysis at different locations on the grid. Regression-based methods \cite{saltelli2000sensitivity} construct a linear regression model between the inputs and the output using input-output pairs generated randomly from the model. The sensitivity measures are determined from the regression coefficients (in standardized form). Regression-based methods generally measure the linear relationship between the input and output, and they are easy to apply, but have limited application for highly nonlinear models. Global and variance-based methods \cite{homma1996importance, saltelli2008global} decompose the output variance into portions attributable to individual input parameters and groups of parameters. These methods can be applied to nonlinear problems and their sensitivity indices indicate how much of the total variance in the output can be attributed to each input parameter (or group of parameters). Although variance-based methods reveal many information about the model, they are more complicated and expensive to apply compared to other methods. 

In the area of renewable energy, a number of studies highlight the importance of SA and UQ when analyzing energy systems. A recently published study \cite{wee2018uncertainty} employed the polynomial chaos expansions (PCE) approach  to perform uncertainty and sensitivity analysis on the maximum power output of a thermoelectric generator (TEG). The uncertainty in the temperature-dependent material properties such as thermal conductivity and electrical resistivity is propagated through the constructed PCE model. The PCE-based uncertainty is compared to the uncertainty calculated by a direct Monte Carlo approach, and some discrepancy between the results was observed. A Monte Carlo approach was used by \cite{tran2018incorporating} to perform global sensitivity analysis on the traditional (e.g. fossil, nuclear) and renewable energy sources under input parameter variability. The main response analyzed is the levelized cost of electricity, while input parameters represent capital cost, operating and maintenance costs, and power generation. A major conclusion from the study is that renewable energy is subjected to high levels of uncertainties in both technical and economic performance due to their initial stages of development \cite{tran2018incorporating}. Next, a regression model was created to evaluate the performance of solar concentrated TEG \cite{rehman2017performance}. Represented inputs to the model include natural, design, and operational parameters of the system including the TEG intrinsic material properties, heat transfer coefficients, solar flux, and others. Least-squares fit of the system output based on data generated over a wide range of the input parameters was used to create the regression model. The study concluded that module area, concentration ratio, solar radiation intensity, and the absorptivity of the receiver are the most influential input parameters to this system \cite{rehman2017performance}. Parameter sensitivity and uncertainty analyses were performed on proton-exchange membrane fuel cells by \cite{mawardi2006effects, min2006parameter,noguer2015pemfc} and a review of SA methods for building energy is conducted by \cite{tian2013review}. 

These UQ and SA methods are highly transferable to many disciplines of engineering, In fact, our previous experience in UQ and SA focused on nuclear reactor engineering where the best estimate plus uncertainty (BEPU) practice is used to replace overly conservative estimates. This BEPU method can lead to greater cost efficiency and a more thorough understanding of the energy system. Previously, a data-driven sampling-based UQ framework was developed to quantify the uncertainties in the reactor kinetic parameters \cite{radaideh2018framework}, which are important for reactor safety. Efforts on global methods and variance decomposition are performed using the Shapley effect and Sobol indices in \cite{radaideh2019shapley,radaideh2018variance,radaideh2018application} with a focus on nuclear data uncertainties. Also various deterministic and stochastic methods have been used to analyze the behaviour of a nuclear spent fuel transportation/storage cask \cite{radaideh2018criticality,radaideh2018onuncertainty,radaideh2019advanced,price2019advanced}. In addition, an integrated framework based on Bayesian statistics has been developed to quantify various uncertainty types in engineering models with a focus on nuclear thermal-hydraulics models \cite{radaideh2019integrated}. As mentioned earlier, since most of these methods have generic and non-intrusive features, they are extendable to other areas of engineering. Although BEPU practice for renewable energy is not as urgent as for nuclear power in terms of safety, it is still important to assess system efficiency and cost benefits of any proposed model with a high degree of understanding. As concluded by \cite{tran2018incorporating}, renewable and sustainable energy sources suffer from large uncertainties both technically and economically. After UQ, system performance can be determined with confidence bounds to asses the system's economic feasibility. The major issue of performing SA and UQ is its expense and complexity (depending on the method used). Consequently, adaptation and application of various SA and UQ methods to energy systems are presented in this paper. The methods are implemented in an integrated and automated form to perform SA and UQ of any energy system (in analytic form in this paper) which makes SA and UQ results a by-product of the system after being modeled. The methods are described and demonstrated using benchmark functions to demonstrate their advantages/disadvantages. Afterward, a practical test on power production from a molten carbonate fuel cell (MCFC) is conducted, as MCFC sensitivity and uncertainty have not been highlighted in previous studies. 

The remaining sections of this paper are organized as follows: section \ref{sec:method} describes various methods to perform SA and UQ both in deterministic and stochastic approaches. Section \ref{sec:test_models} describes the test models used for numerical tests (e.g. Ishigami and Sobol functions) and practical tests (e.g. MCFC), at which the prescribed methods are applied. Section \ref{sec:tests} presents results of applying the SA and UQ methods on benchmark and simple functions. Section \ref{sec:mcfc_res} presents the results of SA and UQ on MCFC power output and efficiency. The conclusions of this study are presented in section \ref{sec:summary}. 

\section{Methodology}
\label{sec:method}

Before starting the description of the proposed methods, basic terminology and definitions used in this section should be defined:
\begin{itemize}
    \item The inputs to the model will be referred to as input parameters or simply parameters, and $x_i$ is used for their notation. A model output will be referred to as a response of interest (RoI) or output, and $y$ is used to refer to a generic RoI.
    \item The number of input parameters to the model, which is known as problem dimensionality, is expressed by $n_x$, and the number of model outputs is given by $n_y$.
    \item A quantity is shown to represent a vector if it is presented with an arrow accent (e.g. $\vec{X}$). Boldface is used if a variable represents a matrix (e.g. $\bm{X}$).
\end{itemize}

\subsection{Sensitivity Analysis}

Four different SA methods are discussed in the following subsections: (1) local SA using linear perturbations or one-at-a-time, (2) standardized regression coefficients, (3) partial correlation coefficients, and (4) Morris screening.   

\subsubsection{One-at-a-time (OAT)}
\label{sec:oat}

Local SA methods investigate how the RoI behaves when changing each input parameter individually. Linear perturbation theory or OAT is the most common local method to perform local SA. The sensitivity index for each input parameter can be calculated by evaluating the partial derivatives of a RoI ($y$) with respect to each input parameter ($x_i$) as 
\begin{equation}
\label{eq:S_local}
    S_i^{OAT} = \frac{\partial y}{\partial x_i}, \quad i=1,...,n_x,
\end{equation}
where $i$ is the input parameter index, $y$ is a single RoI, $x_i$ is the $i^{th}$ input parameter. In most cases, analytical solutions of these partial derivatives do not exsist or are prohibitively difficult to find especially if the response is calculated by complex and nonlinear formulas. In this case, sensitivity index can be determined numerically using first-order finite difference
\begin{equation}
  f'(x) = \frac{f(x+h)-f(x)}{h} + O(h), 
\end{equation}
or 
\begin{equation}
  S_i^{OAT} = \frac{y'-y_0}{x'_i-x_{i_0}}, 
\end{equation}
where $y'$ is the RoI calculated at the perturbed input $x'_i = x_i + \delta x_i$, and $y_0$ and $x_{i_0}$ are the nominal values for the RoI and the $i^{th}$ input parameter, respectively. It is common for the sensitivity index to be normalized to ensure its dimensionless from as follows 

\begin{equation}
\label{eq:oat_norm}
  S_{i,norm}^{OAT} = \frac{(y'-y_0)/y_0}{(x'_i-x_{i0})/x_{i_0}}.  
\end{equation}

It is common to use $S_{i,norm}^{OAT}$ for the importance ranking of the parameters to isolate the parameter unit from affecting the ranking, while $S_i^{OAT}$ is used for deterministic uncertainty propagation as will be described later in this section. The perturbation factor $\delta x$ should be carefully selected by the analyst. However, it is important to ensure that the perturbation is physical and falls in the input parameter range, and not too small to avoid polluting the $S_i^{OAT}$ estimation with computer finite precision residuals. The computational cost of OAT method is $n_x+1$ model evaluations, where the additional evaluation is to calculate the nominal value of the output ($y_0$). To achieve higher accuracy, central finite difference can be used instead, where the model is evaluated at both $x_i + \delta x_i$ and $x_i - \delta x_i$. This implies that the computational cost is doubled for central finite difference. 

\subsubsection{Standardized Regression Coefficients (SRC)}
\label{sec:src}

Regression-based methods measure the strength of the linear relationship between the input parameters and the RoI. Sensitivity measures can be found by analyzing a large number of random samples as input-output pairs. This class of methods is suitable when the response varies linearly with the input parameters. The advantages of regression-based methods are their capability to be measured directly by post-processing the Monte Carlo samples. The method of Standardized regression coefficients (SRC) is described in this subsection. SRC fits a linear model between input parameters and the output. Sensitivity measures are determined by calculating the standardized regression coefficients of the linear model for each parameter. SRC is recommended when the coefficient of determination ($R^2$) of the linear model is high. Moving on, a variation of SRC called standardized rank regression coefficients (SRRC) can also be used where the input and output variables are replaced by their ranks, and the regression process is performed over the ranks. SRRC is preferred when the $R^2$ of the linear model is low. This occurs when dealing with nonlinearity or outlier data points \cite{volkova2008global}. Assume that the model output can be related to the input using a general linear model which fits the output ($y$) with the independent input parameters ($x_j$) as
\begin{equation}
y_i = \alpha_0 + \sum_{j=1}^{n_x} \alpha_j x_{ij} + \epsilon_i, \qquad i = 1, 2, \ldots, n_s,
\label{eq:mlr_src}
\end{equation}
where $\epsilon_i \sim N(0, \sigma^2)$ is the residual, coefficients $\alpha_j$ are the ordinary regression coefficients usually determined by least-squares. Finally, $n_s$ is the number of data points (e.g. random samples) used to construct the model. The data points here represent random samples of the input parameters and their corresponding calculated output. The value of the coefficients $\alpha_j$ depends on the units of the input parameter $x_j$. This leads to a difficulty in using $\alpha_j$ for importance or sensitivity ranking. A common practice is to rescale both the response and input parameters by their mean and variance according to the following:
\begin{equation}
    x_j^* = \frac{x_j - \bar{x}_j}{\sigma_{x_{j}}}, \qquad y^* = \frac{y - \bar{y}}{\sigma_{y}},
\end{equation}
hence, the regression model in Eq.\eqref{eq:mlr_src} can be rewritten in standardized form as 
\begin{equation}
y_i^* = \sum_{j=1}^{n_x} \alpha_j^* x_{ij}^* + \epsilon_i, \qquad i = 1, 2, \ldots, n_s,
\label{eq:mlr_scaled}
\end{equation}
where $\alpha_j^*, (j=1,..,n_x)$ are the model standardized regression coefficients, and they express the linear strength of each input parameter $x_j$ independent of its unit. SRC/SRRC methods measure the change in the response per unit change in an input, when all other inputs remain fixed.  

\subsubsection{Partial Correlation Coefficients (PCC)}
\label{sec:pcc}

Partial Correlation Coefficient (PCC) is another method to perform SA on a model. Given the regression model in Eq. \eqref{eq:mlr_src}, PCC can be defined as a measure of the strength of linear correlation between an input parameter and the response when the other input parameters remain unchanged. Similar to SRRC, partial rank correlation coefficients (PRCC) can be used instead of PCC, if analysis on the input and output ranks is performed (e.g. for nonlinear models). PCC can be used to rank the input parameters based on their linear correlation (e.g. Pearson, Spearman) with the response. Lets assume we have a system with response $y$ and two input parameters $x_1, x_2$,  PCC between the response $y$ and $x_1$ given $x_2$ is fixed is 

\begin{equation}
  \rho_{y,x_1 | x_2} = \frac{\rho_{x_1,y}-\rho_{x_1,x_2}\rho_{x_2,y}}{\sqrt{(1-\rho_{x_1,x_2}^2)(1-\rho_{x_2,y}^2)}}, 
\end{equation}
where the correlation coefficient between any two general variables ($u, v$) can be defined as
\begin{equation}
    \rho_{u,v} = \frac{\sum_i (u_i-\bar{u})(v_i-\bar{v})}{\sqrt{\sum_i (u_i-\bar{u})^2}\sqrt{\sum_i (v_i-\bar{v})^2}},
\end{equation}
where $\bar{u}$ and $\bar{v}$ are the statistical mean of the variables $u$ and $v$, respectively, and $i$ is the sample index. PCC can provide more global information about the linear relationship between the input and the output, and can account for any correlation between the input parameters. The fundamental difference between SRC and PCC is that SRC does not consider that a correlation between $x_i$ and $y$ can be a consequence of a second parameter’s influence ($x_j$) due to the interaction effect. This means that a second parameter's influence is included in SRC measure \cite{volkova2008global}. However, PCC measures the correlation between $x_i$ and $y$ while fixing other input parameters to exclude their influence \cite{saltelli2000sensitivity}. In general, for uncorrelated inputs, SRC and PCC importance ranking is expected to be identical. For highly correlated inputs, differences between SRC and PCC may appear, and PCC is more preferred. For models that are both nonlinear and non-monotonic, global variance decomposition with Sobol indices is preferred, which is discussed later in this section.   

\subsubsection{Morris Screening}
\label{sec:morris_theory}

Morris \cite{morris1991factorial} introduced one of the premier methods for input parameter screening. This method is the global version of OAT method. The Morris method relies on computing several incremental ratios (also called elementary effects) for each input parameter, which are then averaged to assess the overall importance of the input parameter. Qualitatively, elementary effects are similar to the OAT sensitivity in terms of their calculation, with exception that elementary effects are more global in nature. With these importances, the Morris method can be used for dimensionality reduction. Morris screening has the advantage of efficiency and moderate cost compared to other global methods which need large number of samples. The Morris method follows the following major steps \cite{morris1991factorial,campolongo2007effective}:

\begin{itemize}
    \item A base vector ($\vec{x}_0$) of the input parameters is generated randomly from the support of each input parameter. 
    \item An $n_x$-dimensional grid with $p$ levels is constructed. One of the parameters in the base vector is perturbed based on the grid spacing of that parameter to form a new vector. The new vector becomes the base to perturb the next parameter. The process is repeated until all parameters in the system are perturbed.   
    \item To calculate statistics for the elementary effects, the experiment is repeated for $R$ trajectories with a random initial base vector for each trajectory. At each trajectory, $n_x+1$ points are generated under the restriction of OAT perturbation between two successive experiments. To explain, in trajectory $R_2$, experiment 4 has one input perturbed compared to experiment 3, but has more than one perturbation compared to experiment 1. This leaves the total cost of Morris experiment to be $R(n_x+1)$.
\end{itemize}

The elementary effect for input parameter $i$ is defined as 
\begin{equation}
    d_i^{(r)} = \frac{f(x_i^{(r)})-f(x_{i-1}^{(r)})}{\Delta_i}, \quad i=1,...,n_x,
\end{equation}
where $x_i^{(r)}= x_{i-1}^{(r)}  + \Delta_ie_i$. The $\Delta_i$ is grid spacing for input $i$ and it is proportional to $1/(p-1)$, and $e_i$ is a vector of zeros except for the parameter $i$ at which the value is unity. According to \cite{campolongo2007effective}, number of levels $p$ is recommended to be an even number and grid spacing is recommended to be $\Delta = \frac{p}{2(p-1)}$. 

After calculating the elementary effects based on the samples in all trajectories (i.e. design of experiments), they can be processed to calculate their statistics such as the mean (which measures the input parameter importance)
\begin{equation}
\label{eq:dmean}
    \mu_i = \frac{1}{R} \sum_{r=1}^{R} d_i^{(r)}, \quad i=1,...,n_x.
\end{equation}

To avoid the effect of positive/negative indices' cancellation, which could occur in non-monotonic functions, the mean can be calculated by the absolute value of the elementary effects as proposed by \cite{campolongo2007effective} 
\begin{equation}
\label{eq:dmean}
    \mu_i^* = \frac{1}{R} \sum_{r=1}^{R} |d_i^{(r)}|, \quad i=1,...,n_x.
\end{equation}

The variance of the elementary effects can be calculated by 
\begin{equation}
\label{eq:dsd}
    \sigma_{i}^2 = \frac{1}{R-1} \sum_{r=1}^{R} (d_i^{(r)}-\mu_i)^2 , \quad i=1,...,n_x,
\end{equation}
which provides information about the nonlinearity in the model and/or the interactions between the input parameters. Morris \cite{morris1991factorial} mentioned that his method cannot distinguish between the contribution of nonlinearties and interactions. The previous statistics can be used to rank and classify the input parameters into three main categories \cite{pujol2009simplex}:

\begin{itemize}
    \item Inputs with insignificant effect (small $\mu_i^*$)
    \item Inputs with significant effect (large $\mu_i^*$), and insignificant nonlinearity and interactions (small $\sigma_{i}^2$).
    \item Inputs with significant effect (large $\mu_i^*$), and significant nonlinearity and/or interactions (large $\sigma_{i}^2$).
\end{itemize}

Finally, it is worth mentioning that Morris elementary effects should be normalized similar to OAT Eq.\eqref{eq:oat_norm}, if it will be used for importance ranking of parameters with different units. In this study, the parameters $\mu_{i,norm}$ and $\mu_{i,norm}^{*}$ are to refer to the normalized version of the means $\mu_{i}$ and $\mu_{i}^{*}$, respectively.  

\subsection{Uncertainty Quantification}

Two main methods are described to perform forward UQ: (1) deterministic-based UQ using uncertainty propagation rules and (2) sampling-based UQ using Monte Carlo methods.

\subsubsection{Deterministic-based UQ}
\label{sec:det_uq}

The output uncertainty can be calculated deterministically by uncertainty propagation rules through combining the parametric sensitivity profiles with the variance-covariance matrix of the parameters. In matrix form, the sensitivity matrix contains the sensitivity coefficients of each parameter with respect to the RoIs. This matrix can be calculated using OAT presented in section \ref{sec:oat} or using the Morris method described in section \ref{sec:morris_theory}. The variance-covariance matrix contains the uncertainty information of the parameters with the variance of each parameter on the diagonal, and the covariance between the parameters on off-diagonal entries. The uncertainty in the RoI can be written as 
\begin{equation}
\label{eq:uqq}
    \underbrace{C_y}_\text{$n_y \times n_y$} = \underbrace{S}_\text{$n_y \times n_x$} \underbrace{C_x}_\text{$n_x \times n_x$} \underbrace{S^T}_\text{$n_x \times n_y$},
\end{equation}
where $C_x$ and $C_y$ are the variance-covariance matrices of the input parameters and RoIs, respectively, and $S$ is the sensitivity matrix of the input parameters. For simplicity, consider a single response ($n_y=1$), the sensitivity matrix becomes a vector 
\[
S=
\begin{bmatrix}
S_{x_1}  & S_{x_2} & \cdots & S_{x_{n_x}}  \\
\end{bmatrix}
,\]
and the variance-covariance matrix remains unaffected as follows
\[
C_x=
\begin{bmatrix}
\sigma_{x_1}^2 & COV(x_1,x_2) & \cdots &  COV(x_1,x_{n_x}) \\
COV(x_2,x_1) & \sigma_{x_2}^2 & \cdots &  COV(x_2,x_{n_x})\\
\vdots & \vdots & \ddots & \vdots \\
COV(x_{n_x},x_1) & COV(x_{n_x},x_2) & \cdots & \sigma_{x_{n_x}}^2\\
\end{bmatrix}
.\]
In this case, $C_y$ is a scalar quantity which represents the variance in a single RoI, and $S$ becomes a vector contains the sensitivity coefficients between the RoI and each of the parameters. If the input parameters are uncorrelated, the off-diagonal entries of $C_x$ are zero. In this case, Eq.\eqref{eq:uqq} becomes similar to the propagation formula commonly used by experimentalists to perform uncertainty propagation \cite{ku1966notes}.   

\subsubsection{Monte Carlo Sampling-based UQ}
\label{sec:mc_uq}

The Monte Carlo approach requires four primary steps: (1) assigning random distributions to the input parameters (e.g. normal, beta, lognormal, etc.), (2) generating $n_s$ global random samples of the parameters from the assigned distributions, (3) propagating the random samples through the model, and (4) calculating the statistical moments of the RoI (e.g. mean, variance, etc.).  First, lets assume for demonstration that all input parameters follow an uncorrelated univariate normal distribution, each input parameter can be sampled as follows   

\begin{equation}
    x^{(j)}_i \sim N(\mu_{x_i},\sigma_{x_i}^2) \quad j =1,.., n_{s},
\end{equation}
where $\mu_{x_i}$ and $\sigma_{x_i}^2$ are the mean and variance of the parameter $x_i$, which can be obtained from experiments or expert judgment. The random samples $x^{(j)}_i$ for $j =1,.., n_{s}$ are then passed through the model (e.g. analytic, computer code, etc.), and the RoI is calculated for each random sample set.  Now, the first and second statistical moments of the RoI can be calculated using 
\begin{equation}
\label{eq:ymean}
    \overline{y} = \frac{1}{n_{s}} \sum_{j=1}^{n_{s}} y^{(j)},
\end{equation}
\begin{equation}
\label{eq:ysd}
    \sigma_y^2 = \frac{1}{n_{s}-1} \sum_{j=1}^{n_{s}} (y^{(j)}-\overline{y})^2.
\end{equation}

The 95\% confidence interval (CI) for the RoI can be calcualted using
\begin{equation}
\label{eq:yconf}
    95\% \text{ CI} = \overline{y} \pm 1.96 \frac{\sigma_y^2}{\sqrt{n_s}}.
\end{equation}

\subsection{Variance Decomposition with Sobol Indices}
\label{sec:sobol}

Variance-based or global methods \cite {saltelli2008global} decompose the response variance into portions attributable to individual input parameters and groups of parameters. These methods yield sensitivity indices that express how much of the total variance in the output can be attributed to each input parameter. Variance-based methods rely on the probabilistic nature of the input-output. Primary advantages of these global methods over local methods are their ability to account for interactions and nonlinearity between the input parameters and the output. The Sobol study \cite{sobol1993sensitivity} was one of the earliest efforts on variance decomposition under the condition of independent/uncorrelated input parameters. He proposed decomposing the output variance ($Var[y]$) into portions attributable to each input parameter over the input range/support. The first order effect can be defined as
\begin{equation}
\label{eq:V_i}
 V_i = Var[E[y|x_i]] = Var [y] - E[Var[y|x_i]],   
\end{equation}
where $V_i$ represents the reduction in $Var[y]$ when $x_i$ is fixed. The total effect is the complement of the first order effect
\begin{equation}
\label{eq:T_i}
 V_i^T = Var [y] - Var[E[y|\vec{x}_{\sim i}]],   
\end{equation}
which expresses the remaining variance of $y$ when all input parameters other than $x_i$ (i.e. $\vec{x}_{\sim i}$) are fixed. The Sobol decomposition is defined by 
\begin{multline}
\label{eq:decomp}
    F(x_1,...,x_d) = F_0 + \sum_{i=1}^d F_i(x_i) + \sum_{1 \leq i < j \leq d} F_{ij} (x_i,x_j) \\ + ... + F_{12...d} (x_1, x_2, ..., x_d),
\end{multline}
where two conditions should be held: (1) $F_0$ is constant and equal to the expected value of $F(\vec{x})$, (2) all terms in the functional decomposition are orthogonal (i.e. integral of the sum with respect to its own variables is zero)
\begin{equation}
    \label{eq:cond2}
    \int_{x_d} F_{{i_1}, ..., i_s} (x_{i_1},...,x_{i_s}) d\vec{x}_{i_k} = 0, \qquad 1 \leq k \leq s.
\end{equation}

The decomposition in Eq.\eqref{eq:decomp} under the prescribed conditions is unique according to \cite{sobol1993sensitivity}. Consequently, the terms in Eq.\eqref{eq:decomp} can be defined over the parameter support ($x_d$) as follows

\begin{equation}
    \label{eq:f0}
    F_0 = \int_{x_d} F(x_1,...x_d) dx_1...x_d,
\end{equation}

\begin{equation}
    F_i(x_i) = \int_{x_d} F(x_1,...x_d)d\vec{x}_{\sim i} - F_0,
\end{equation}

\begin{equation}
    \label{eq:fij}
    F_{ij}(x_i,x_j) = \int_{x_d} F(x_1,...x_d)d\vec{x}_{\sim ij} - F_0 -F_i(x_i) - F_j(x_j),
\end{equation}
and so on for the higher order terms. For square integrable $F(\vec{x})$, the total variance is defined as 

\begin{equation}
\label{eq:var}
    V = Var[y] = \int_{x_d} F^2(\vec{x}) d\vec{x} - F_0^2,
\end{equation}
and in a similar manner, partial variances can be computed by 
\begin{multline}
   V_{i_1, ..., i_s} = \int_{x_d} F_{i_1,...,i_s}^2 (x_{i_1},...,x_{i_s}) dx_{i_1}...dx_{i_s} \\ \qquad 1 \leq i_1 < ... < i_s \leq d, \quad s = 1,...,d.  
\end{multline}

The first order Sobol index in normalized form is defined by  the ratio of the partial variance to the total variance
\begin{equation}
    S_{i_1, ..., i_s} = \frac{V_{i_1,...,i_s}}{V}.
\end{equation}

The first order Sobol index $S_i$ represents the effect of each input parameter alone on the total variance. The second order index $S_{ij}$ represents the effect of the interactions between the $i^{th}$ and the $j^{th}$ parameters. The third and higher order terms can be interpreted similarly. Since Sobol indices represent fractions of the total variance, the first order index should sum to 1 as follows
\begin{equation}
\label{eq:sj_cond}
    \sum_{i=1}^d S_i + \sum_{1 \leq i\leq j < d} S_{ij} + ... + S_{12...d} = 1.
\end{equation}

The total effect $T_i$ for each parameter, which represents its contribution alone as well as its interactions with all other parameters, is defined as 
\begin{equation}
\label{eq:STi}
  T_i = \frac{E[Var[y|x_{\sim i}]]}{Var[y]}  = 1- \frac{Var[E[y|x_{\sim i}]]}{Var[y]}= 1 - S_{\sim i}.   
\end{equation}

Two general conditions for the first and total effects should be satisfied
\begin{equation}
\label{eq:cond1}
     \sum_{i=1}^d T_i \geq 1 \geq \Bigg[\sum_{i=1}^d S_i + \sum_{1 \leq i\leq j < d} S_{ij} + ... + S_{12...d}\Bigg], 
\end{equation}
and
\begin{equation}
\label{eq:cond2}
   T_i \geq S_i.
\end{equation}    
where the equality holds when the parameter $i$ has no interactions with other parameters. 

Analytical calculations of global sensitivity indices by evaluating the integrals in Eqs.\eqref{eq:f0}-\eqref{eq:var} are usually prohibitively difficult for complex mathematical/computer models. Alternatively, Monte Carlo methods are used to calculate Sobol indices. The reliance on Monte Carlo sampling leads to the main disadvantage of this approach, since the calculations can involve a large number of model runs to ensure sampling convergence of the indices. A method for Monte Carlo estimation of Sobol indices based on \cite{glen2012estimating} is considered in this study. Glen and Issac \cite{glen2012estimating} evaluated the performance of different correlation techniques to estimate Sobol indices. They also applied correction factors to reduce the effect of spurious correlation in the Sobol estimation. Based on their study \cite{glen2012estimating}, the authors confirmed that the technique ``D3" in Table 1 in their paper yielded the best performance and results. This D3 technique was verified and applied previously in \cite{radaideh2019shapley,radaideh2018variance,radaideh2018application}. A step-by-step of the ``D3" algorithm is shown as Algorithm \ref{alg:sobolglen} as reproduced from \cite{radaideh2019shapley}. The inputs to the algorithm are:  (1) two independent matrices ($\bm{X}$,$\bm{X'}$) sampled from the joint/marginal distribution of the input parameters and (2) the model being evaluated (e.g. analytic function, computer code, etc.). The size of each matrix (i.e. $\bm{X}$,$\bm{X'}$) is $N \times n_x$, where the columns represent the input parameters, and the rows represent their random samples. The cost for this algorithm is $N(2+2n_x)$ model evaluations. 

\begin{algorithm}[h!]
\small
\caption{\small Sobol Indices Calculation with Spurious Correction \cite{glen2012estimating,radaideh2019shapley,radaideh2018variance}}
\label{alg:sobolglen}
\begin{algorithmic}
  \State \textbf{\%Comment: (1) Prepare data}
  \State (a) Set $N$ (number of samples) and $n_x$ (number of input parameters). 
  \State (b) Sample $\bm{X}$, $\bm{X'}$ matrices (size $N \times n_x$) independently from $\bm{G}_{X}$ (e.g. normal, uniform).
  \State \textbf{\%Comment: (2) Radial Sampling}
  \State (a) Evaluate $\vec{g}_0 = F(\bm{X}), \vec{g}_0'=F(\bm{X'})$
  \For {$i = 1$:$N$} 
         \For{$j = 1$:$d$}
             \State (b) Define $\vec{x}= \bm{X}_{i,:}$, $\vec{x}'=\bm{X'}_{i,:}$
             \State (c) Set $\vec{x}_j=\bm{X}'_{i,j}$, $\vec{x}'_j=\bm{X}_{i,j}$ 
             \State (d) Evaluate $\bm{g}_{i,j} =F(\vec{x})$,  $\bm{g}'_{i,j} =F(\vec{x}')$. 
        \EndFor
  \EndFor 
  \State \textbf{\%Comment: (3) Standardization}
  \State (a) Calculate the sample mean and standard deviation (i.e. column-wise for matrices) for $\vec{g}_0$, $\vec{g}'_0$, $\bm{g}$, $\bm{g}'$.
  \State (b) Replace $\vec{g}_0 = (\vec{g}_0-\bar{g}_0)/\sigma_{g_0}$,     $\vec{g}_0' = (\vec{g}_0'-\bar{g}_0')/\sigma_{g_0'}$ 
  \State (c) Replace $\bm{g} = (\bm{g}-\vec{\bar{g}})/\vec{\sigma}_{g}$,     ${\bm{g}}' = (\bm{g}'-\vec{{\bar{g}}}')/\vec{\sigma}_{{g}'}$
  \State \textbf{\%Comment: (4) Calculate the Parameters with Spurious Correction}
  \State (a) Set $\vec{C}=0,\vec{C}'=0,\overrightarrow{CF}=0$
   \For {$i = 1$:$N$} 
        \For{$j = 1$:$d$} 
            \State (b) $\vec{C}_j = \vec{g}_{0_i} \times \bm{g}_{i,j}' + \vec{g}'_{0_i} \times \bm{g}_{i,j} + \vec{C}_j$   
            \State (c) $\vec{C}'_j = \vec{g}_{0_i} \times \bm{g}_{i,j} + \vec{g}'_{0_i} \times \bm{g}_{i,j}' + \vec{C}'_j$ 
            \State (d) $\overrightarrow{CF}_j = \vec{g}_{0_i} \times \vec{g}_{0_i}  + \bm{g}_{i,j} \times \bm{g}_{i,j}' + \overrightarrow{CF}_j$
        \EndFor
   \EndFor
   \State (e) Set $\vec{C}=\vec{C}/(2N), \vec{C}'=\vec{C}'/2N, \overrightarrow{CF}=\overrightarrow{CF}/2N$.
   \State (f) Calculate $\vec{E}=\frac{(\vec{C}- \overrightarrow{CF} \times \vec{C}')}{1- \overrightarrow{CF} \times \overrightarrow{CF}}$, $\vec{E}'=\frac{(\vec{C}'- \overrightarrow{CF} \times \vec{C})}{1- \overrightarrow{CF} \times \overrightarrow{CF}}$
   \State \textbf{\%Comment: (5) Calculate Sobol indices $S_j, T_j$, $(j=1,..,d)$}
   \State (a) $\vec{S}=\vec{C}- \frac{\overrightarrow{CF} \times \vec{E}'}{1-\vec{E} \times \vec{E'}}$
   \State (b) $\vec{T}= 1- \vec{C'} + \frac{\overrightarrow{CF} \times \vec{E}}{1-\vec{E} \times \vec{E'}}$
\end{algorithmic}
\end{algorithm}

\section{Test Models}
\label{sec:test_models}

In this section, three selected models are used as benchmarks to demonstrate SA and UQ methods: (1) the Sobol function, (2) the Morris function, and (3) the Ishigami function. Furthermore, a special set of simple functions are created to analyze model parameter interaction and nonlinearity. 

\subsection{Benchmark Functions}

Sobol's g-function \cite{saltelli2000sensitivity} has $n_x$ inputs, all drawn from a uniform distribution between [0,1]. The function is preferred for sensitivity and uncertainty analysis due to its fair complexity and its analytical solution for Sobol indices. The function is expressed as follows
\begin{equation}
\label{eq:sobol_fun}
    f_{Sobol}(\vec{x}) = \prod_{i=1}^{n_x} \frac{|4x_i-2|+a_i}{1+a_i}, \quad i=1,2, \cdots,n_x,
\end{equation}
the constant $a_i \geq 0$ can be selected by the analyst. In this study, $n_x=8$ is fixed and $a_i=(0,1,2,3,5,10,20,50)$ is used. The reason for this selection will be demonstrated later in the next section as $a_i$ value controls the parameter's importance. 

Moving forward, the Morris function \cite{saltelli2000sensitivity,morris1991factorial} has 20 inputs, all drawn from a uniform distribution between [0,1]. The function has the feature of high-dimensionality and non-monotonicity, and it was created by the author \cite{morris1991factorial} to test Morris screening method. The function is expressed by    

\begin{multline}
\label{eq:morris_fun}
    f_{Morris}(\vec{x}) = \beta_0 + \sum_{i=1}^{20} \beta_i w_i + \sum_{i<j}^{20} \beta_{i,j} w_i w_j \\
    + \sum_{i<j<l}^{20} \beta_{i,j,l} w_i w_j w_l + \sum_{i<j<l<m}^{20} \beta_{i,j,l,m} w_i w_j w_l w_m,
\end{multline}
where 
\begin{align}
\begin{aligned}
\label{eq:wi}
w_i=2(x_i-0.5), \quad \text{for } i= 1, 2, \cdots ,20 \text{ except } \{3,5,7\}, \\
w_i=2\Bigg(\frac{1.1x_i}{x_i+0.1} - 0.5\Bigg), \quad \text{for } i=\{3,5,7\}.
\end{aligned}
\end{align}

The $\beta$ coefficients are assigned as follows:

\begin{align}
\begin{aligned}
\beta_0 = N(0,1), \\
\beta_{i}= 20, \quad \text{for } i=1, 2 \cdots, 10, \\
\beta_{i} \sim N(0,1), \quad \text{for } i=11, 12, \cdots, 20, \\
\beta_{i,j}= -15, \quad \text{for } i,j=1, 2 \cdots, 6, \\
\beta_{i,j} \sim N(0,1), \quad \text{for } i,j=7, 8, \cdots, 20, \\
\beta_{i,j,l}= -10, \quad \text{for } i,j,l=1, 2 \cdots, 5, \\
\beta_{i,j,l,m} = 5, \quad \text{for } i,j,l,m=1, 2, \cdots, 4, \\
\end{aligned}
\end{align}
where the remainder of the third ($\beta_{i,j,l}$) and fourth ($\beta_{i,j,l,m}$) order coefficients are set to zero. 

Finally, the Ishigami function \cite{ishigami1990importance} has 3 inputs, all drawn from a uniform distribution between [$-\pi,\pi$]. The Ishigami function is widely used for uncertainty and sensitivity analyses due to its strong nonlinearity and nonmonotonicity. The function is expressed by

\begin{equation}
\label{eq:ish_fun}
    f_{Ish}(\vec{x}) = sin(x_1) + a \cdot sin^2(x_2) + b\cdot x_3^4sin(x_1), 
\end{equation}
the recommended values of $a$ and $b$ are: $a = 7$ and $b = 0.1$ as used before by \cite{marrel2009calculations}.

\subsection{Special Function Set (SFS)}
\label{sec:SFS}

This SFS is suggested by the authors to provide simple functions to demonstrate the aforementioned methods. SFS contains 4 functions that vary between linear and nonlinear. Each function takes 3 parameters. The first function is a sum of three linear variables 

\begin{equation}
\label{eq:f1}
    f_1 (\vec{x}) = x_1 + x_2 + x_3,
\end{equation}
the second function contains an interaction term between $x_1$ and $x_2$ as follows 
\begin{equation}
\label{eq:f2}
    f_2 (\vec{x}) = x_1 + x_1x_2 + x_3, 
\end{equation}
the third function contains two nonlinear terms for $x_2$ and $x_3$ as follows 
\begin{equation}
\label{eq:f3}
    f_3 (\vec{x}) = x_1 + x_2^2 + x_3^3, 
\end{equation}
the fourth function contains both nonlinearity and interaction terms as follows 
\begin{equation}
\label{eq:f4}
    f_4 (\vec{x}) = x_1 + x_1x_2^2 + x_3^3, 
\end{equation}

\begin{equation}
\label{eq:f1}
    f_1 (\vec{x}) = x_1 + x_2 + x_3,
\end{equation}
\begin{equation}
\label{eq:f2}
    f_2 (\vec{x}) = x_1 + x_1x_2 + x_3, 
\end{equation}
\begin{equation}
\label{eq:f3}
    f_3 (\vec{x}) = x_1 + x_2^2 + x_3^3, 
\end{equation}
\begin{equation}
\label{eq:f4}
    f_4 (\vec{x}) = x_1 + x_1x_2^2 + x_3^3, 
\end{equation}

The previous functions can be used to assess the prescribed methods on how they handle linear models, parameter interactions, and/or nonlinearties. 

\subsection{Molten Carbonate Fuel Cell}
\label{sec:mcfc}

Hydrogen fuel used in fuel cells forms a viable energy resource for the future. Fuel cell technology converts the chemical energy of the fuel reactants into electricity \cite{sharaf2014overview}. Fuel cells have different constructions depending on the electrolyte type such as proton exchange membrane fuel cell \cite{vishnyakov2006proton}, phosphoric acid fuel cell \cite{chen2015maximum}, MCFC, and others. Fuel cells (FCs) produces electrical current, water, and waste heat. The waste heat can be recovered to achieve additional power output using systems such as gas turbines, TEG, and/or absorption chillers \cite{yang2014performance,zhang2015performance,arsalis2012modeling}. The standalone MCFC system is analyzed here in depeth, while an analysis of the fuel cell hybrid system performance will be analyzed in future studies. The power ($P$) and efficiency ($\eta$) of the MCFC can be written according to the first law of thermodynamics as \cite{zhang2011performance,wu2016performance,wu2017performance}

\begin{equation}
\label{eq:P_fc}
    P=jA(E-U_{an}-U_{cat}-U_{ohm}),
\end{equation}
and
\begin{equation}
\label{eq:Eta_fc}
    \eta=\frac{n_eF}{-\Delta h}(E-U_{an}-U_{cat}-U_{ohm}),
\end{equation}
where $j$ is the operating current density, $A$ is the effective area of the FC, $n_e$ is the number of electrons involved per reaction, $F$ is Faraday's constant, $\Delta h$ is the molar enthalpy change of the electrochemical reaction in the MCFC, and $E$ is the equilibrium potential which is defined as

\begin{equation}
\label{eq:E}
    E= E_{0} +  \frac {R_{gas}T}{n_{e}F}  \ln\Bigg[ \frac  {\big({p_{H_2,an} \big)  \big(p_{O_2,cat}^{0.5}\big)   \big(p_{CO_2,cat} \big)}}{  \big(p_{H_2O,an} \big)  \big(p_{CO_2,an} \big) }\Bigg],
\end{equation}
where 
\begin{equation}
\label{eq:E0}
    E_{0} = \frac {242000-45.8 T}{ n_{e} F},
\end{equation}
where $R_{gas}$ is the universal gas constant, $T$ is the operating temperature, $E_0$ is the standard potential, and $p_{H_2}$ , $p_{O_2}$ , $p_{H_2O}$, and $p_{CO_2}$ are the partial pressures of $H_2$, $O_2$, $H_2O$ and $CO_2$, respectively. The subscripts $cat$ and $an$ refer to the cathode and anode, respectively. The irreversible losses $U_{an}$, $U_{cat}$, and $U_{ohm}$ are the anode overpotential, cathode overpotential, and ohmic overpotential, respectively, which are defined as follows \cite{zhang2011performance}

\begin{equation}
\label{eq:Uan}
    U_{an} =  2.27\times 10^{-9}  j   \exp{ \Bigg( \frac {E_{act,an}}{ R_{gas} T } \Bigg) }  ( p_{H_2,an}^{-0.42} )   ( p_{CO_2,an}^{-0.17} )   ( p_{H_2O,an}^{-1.00} ),  
\end{equation}

\begin{equation}
\label{eq:Ucat}
    U_{cat} =  7.505\times 10^{-10}   j   \exp{  \Bigg( \frac {E_{act,cat}}{ R_{gas} T } \Bigg) }  ( p_{O_2,cat}^{-0.43} )   ( p_{CO_2,cat}^{-0.09} ),  
\end{equation}

\begin{equation}
\label{eq:Uohm}
    U_{ohm} =  0.5\times 10^{-4}   j   \exp{ \Bigg[ 3016  \Big( \frac{1}{T}- \frac{1}{923} \Big)  \Bigg] }, 
\end{equation}
where $E_{act,an}$ and $E_{act,cat}$ are the anode and cathode activation energy, respectively. 

\section{Numerical Tests}
\label{sec:tests}

\begin{figure*}[!h] 
 \centering
  \includegraphics[width=6.0 in]{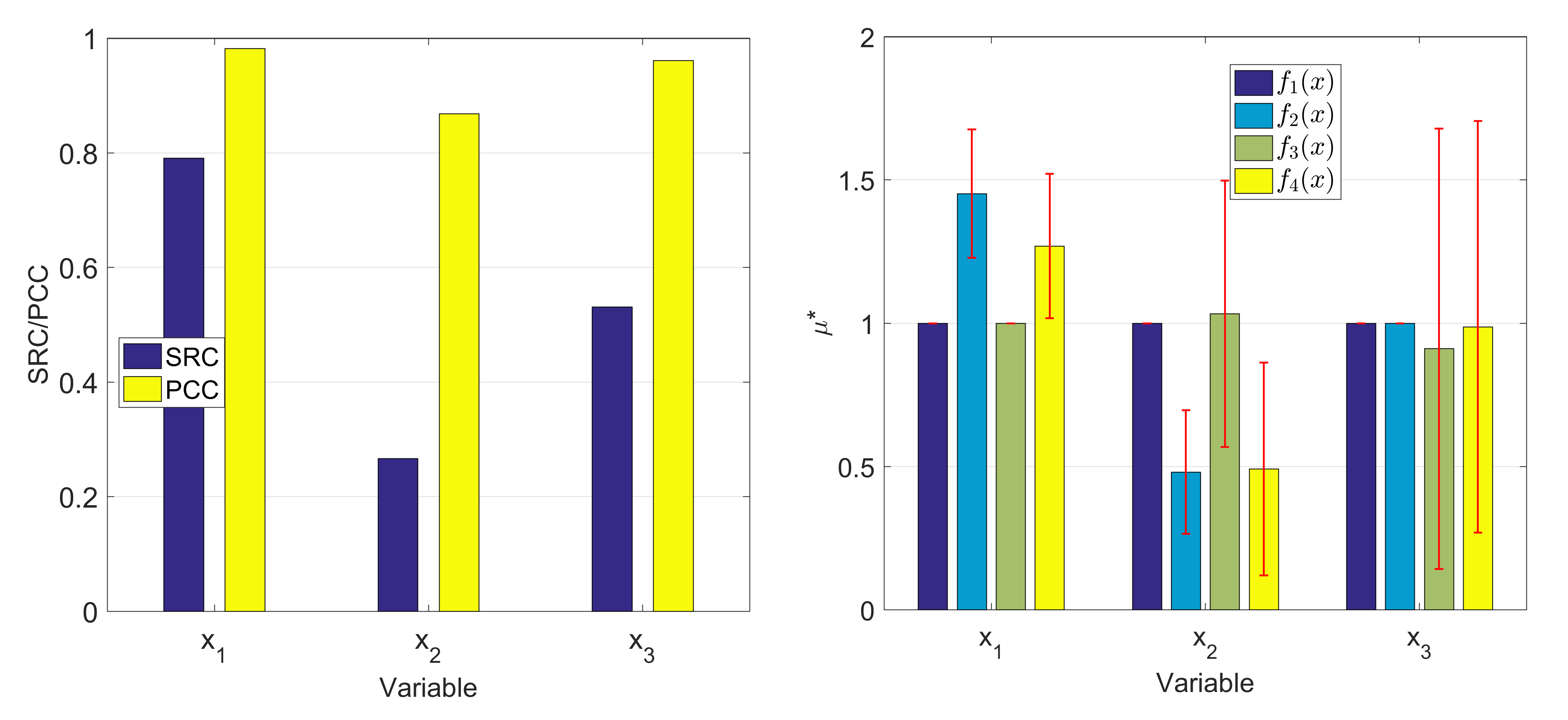}
  \caption{Comparison of SRC and PCC coefficients for the function $f_2(x)$ (left) and comparison of the Morris elementary effects and their standard deviation for the 4 functions in the SFS described in section 3.2 (right)}
  \label{fig:sfs_test}
\end{figure*}

The tests are started by applying the SRC and PCC methods. A comparison of SRC and PCC results is shown in Figure \ref{fig:sfs_test} (left) when applying these methods on $f_2(x)$ in the SFS [See Eq.\eqref{eq:f2}]. The parameters ($x_1,x_2,x_3$) are sampled uniformly from $U[0,1]$ without correlation. Since there is no nonlinearity, the analysis is done on the variables and not on their ranks (i.e. SRRC and PRCC are not used). Again in Figure 1, both SRC and PCC agree on the importance ranking of the three parameters due to the absence of any correlation between the input parameters. The results show that $x_1$ has the largest SRC followed by $x_3$ and finally $x_2$. However, results show that $x_1$, $x_2$ and $x_3$ have similar measures for PCC compared to SRC measures. As mentioned in section \ref{sec:pcc}, SRC includes the interaction between the parameters during the calculations which resulted in giving $x_1$ much larger effect, as $x_1$ contributes individually (the first term in $f_2(x)$) and due to interaction with $x_2$ in the second term. For SRC, $x_2$ has a smaller coefficient as it only contributes by its interactions with $x_1$. Since PCC controls other variables when calculating each correlation coefficient, it gives $x_1$ and $x_3$ close estimates, both slightly higher than $x_2$. To summarize, we expect SRC measures for ($x_1,x_2,x_3$) to be close to each other if SRC is applied to $f_1(x)$. Also, since $f_2(x)$ is additive and monotonic, no negative sensitivity coefficients are observed for any parameter in Figure \ref{fig:sfs_test}(left).   

\begin{figure*}[h!] 
 \centering
  \includegraphics[width=5.5 in]{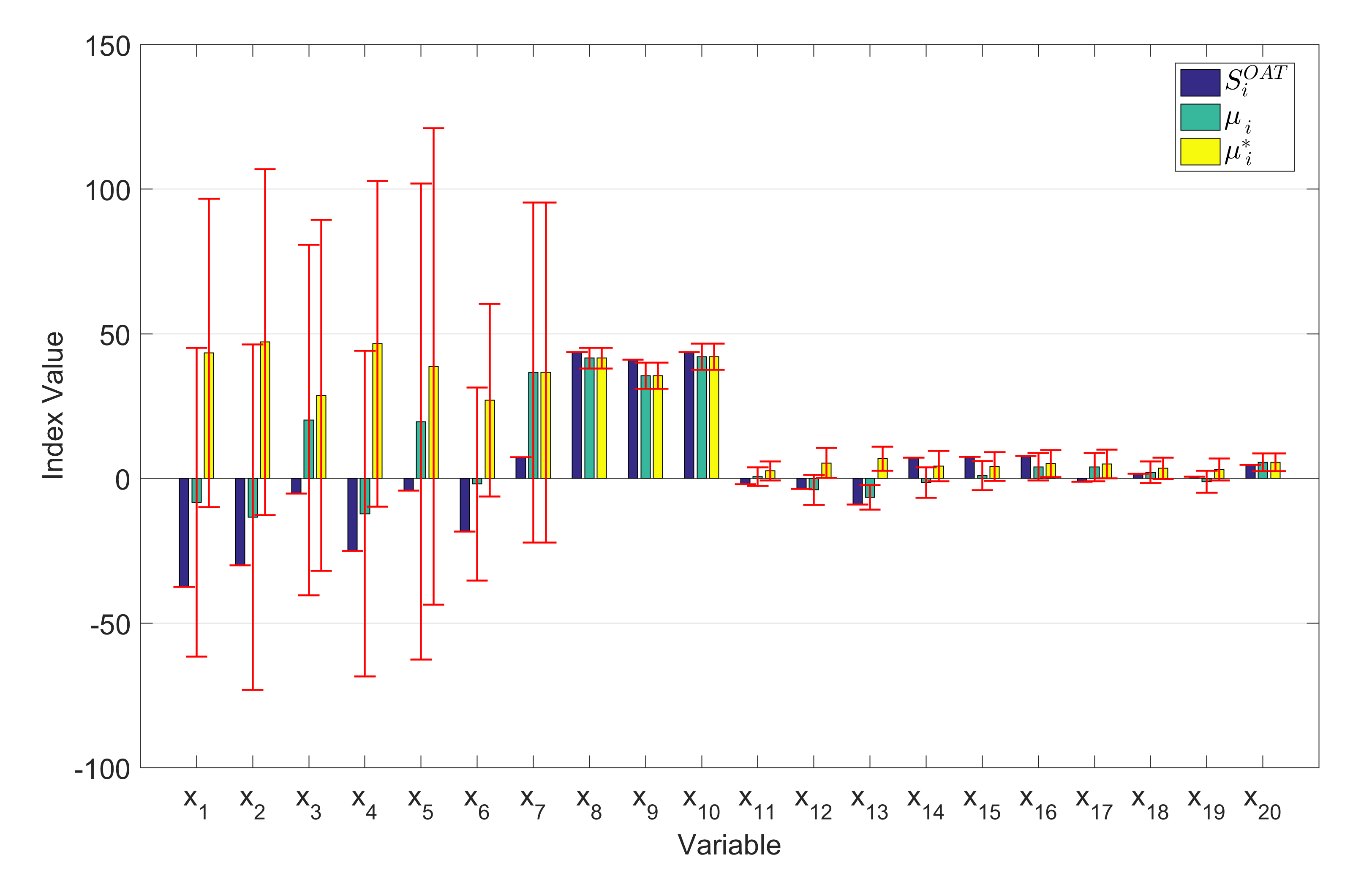}
  \caption{Comparison of the calculated sensitivities of OAT finite difference and Morris methods when applied on Morris function}
  \label{fig:sa_vs_morris_tests}
\end{figure*}

Application of Morris screening on all functions in SFS is given in Figure \ref{fig:sfs_test}(right). The mean of the absolute value of the elementary effects ($\mu^*$) is plotted with its uncertainty. Due to the simplicity and monotonicity of these SFS functions, both $\mu$ and $\mu^*$ have exactly same value (only $\mu^*$ is plotted for clarity). However, these functions demonstrate how Morris screening results in Figure \ref{fig:sfs_test}(right) can be interpreted. First, uncertainty in $x_1$ ($\sigma_{x_1}$) is negligible for $f_1$ and $f_3$, where $x_1$ does not interact with any other parameter, but significant in $f_2$ and $f_4$ where $x_1$ interacts with $x_2$. For $x_2$, interaction with $x_1$ in $f_2$, nonlinearity in $f_3$, and both interaction and nonlinearity in $f_4$ cause large $\sigma_{x_2}$ in these functions. Finally, $x_3$ has only nonlinearity (i.e. third order) in $f_3$ and $f_4$, causing a very large uncertainty in its elementary effects for $f_3$ and $f_4$. These examples show a straightforward demonstration on interpreting Morris results. A more complex application on Morris function with 20 parameters with comparison to OAT finite difference (see section \ref{sec:oat}) is given in Figure \ref{fig:sa_vs_morris_tests}. Morris results are carried out using $R=200$ trajectories, and uniform fine grid spacing of 0.01. General observations from Figure \ref{fig:sa_vs_morris_tests} results are:
\begin{itemize}
\item The presence of four-way interactions in $x_1$-$x_4$ and three-way interactions in $x_1$-$x_5$ cause a large error bar in the elementary effects of these parameters. 
\item The parameters ($x_3, x_5, x_7$) have additional uncertainty contribution from their fractional form as in Eq.\eqref{eq:wi}. 
\item The value of $\mu$ and $\mu^*$ can be significantly different especially for the parameters with large error bars.   
\item OAT sensitivity demonstrates large disagreement with Morris results, especially for the first 7 parameters. This disagreement is due to the significant interaction and nonlinearity for these parameters, which cannot be captured by the local OAT. Fair agreement can be found for other parameters such as $x_8$-$x_{10}$. 
\item The absence of three and four-way interactions ($\beta_{ij,l} = \beta_{ij,l,m} = 0$), the small two-way interaction coefficient [$\beta_{i,j}=N(0,1)$], and small $\beta_i$ diminish the sensitivity of the last ten parameters ($x_{11}$-$x_{20}$), compared to the first ten parameters ($x_{1}$-$x_{10}$).
\end{itemize}

\begin{figure*}[h!] 
 \centering
  \includegraphics[width=5.5 in]{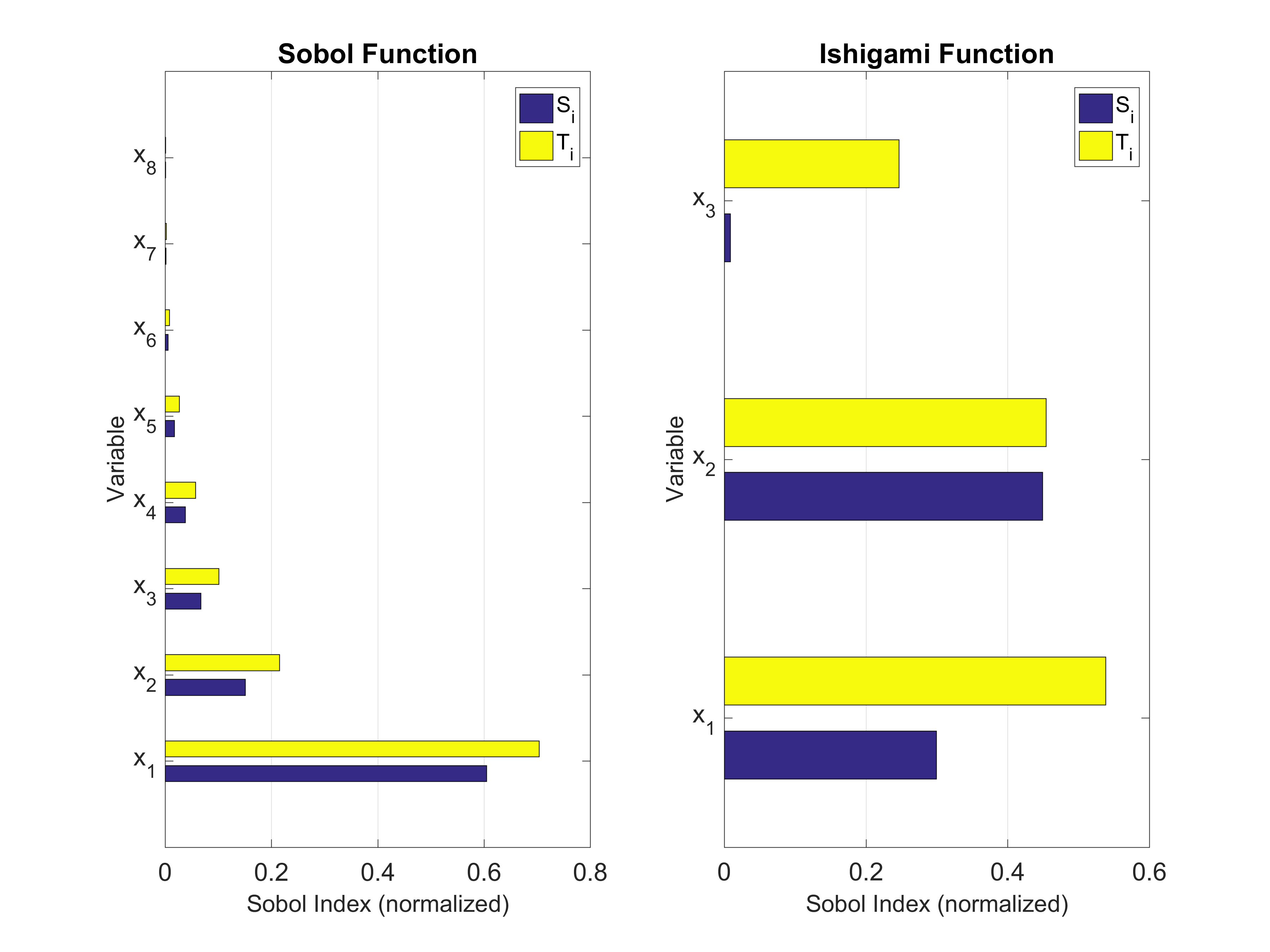}
  \caption{First order ($S_i$) and Total ($T_i$) Sobol indices calculated for two benchmark functions: Sobol function (left) and Ishigami function (right)}
  \label{fig:sobol_tests}
\end{figure*}

Figure \ref{fig:sobol_tests} presents numerical values of the first order ($S_i$) and total ($T_i$) Sobol indices, calculated from Algorithm \ref{alg:sobolglen} for variance decomposition applied to the Sobol function in Eq.\eqref{eq:sobol_fun} and Ishigami function in Eq.\eqref{eq:ish_fun}. First, all 8 input parameters in the Sobol function are sampled independently from a uniform distribution $U[0,1]$. Values of $a_i=\{0,1,2,3,5,10,20,50\}$ are used. It is clear that as $a_i$ increases, the variance contribution of $x_i$ diminishes. In addition, the difference between the total and first-order indices decreases as $a_i$ increases, due to the reduction of the effect of parameter interactions. For the Ishigami function, $x_1$ has a large total index due to its interaction with $x_3$. The parameter $x_2$ has similar first and total indices due to the absence of any interactions with other parameters. However, the parameter $x_3$ has a small first order index, but large total effect due to its interaction with $x_1$. The results of $x_3$ show that the direct effect of the parameter could be negligible, but its indirect effect due to its interactions with other parameters in the system could be significant. In general, the Glen and Issac algorithm \cite{glen2012estimating} provides reliable estimates of Sobol indices as the normalization conditions in Eqs.\eqref{eq:cond1}-\eqref{eq:cond2} are satisfied. Finally, it is worth mentioning that the number of random samples used in these Sobol tests is $10^4$ samples. 

\section{Practical Tests on MCFC}
\label{sec:mcfc_res}

Operating conditions and approximate uncertainties for major MCFC parameters are listed in Table \ref{tab:fc_data}, which are used during the analysis in this section. These uncertainties are obtained from suggestions of previous studies that were based on experimental measurements \cite{placca2009effects,coppo2006influence,basu2006situ,alaefour2012current}. Notice that the input and output uncertainties in this section are expressed mostly in relative form (i.e. $\sigma/\mu\%$), which is dimensionless compared to the absolute uncertainty $\sigma$. This section is divided into two subsections. The first section contains a general UQ analysis to determine the optimum current density ($j^*$) at which the power output is maximized. In this case, the current density is assumed to be deterministic (without uncertainty). In the second subsection, the system at the optimum current $j^*$ is analyzed in terms of its sensitivity and uncertainty, at which 1\% uncertainty is assigned to the optimum current density. Two main RoIs are analyzed from the MCFC: power output ($P$) and system efficiency ($\eta$). 

\begin{table*}[htbp]
  \centering
  \small
  \caption{List of operating parameters used for modeling of the MCFC system with their uncertainty \cite{zhang2011performance,wu2016performance,wu2017performance, placca2009effects,coppo2006influence,basu2006situ,alaefour2012current}}
    \begin{tabular}{ccc}
    \toprule
    Parameter & Nominal Value & Uncertainty  \\
    \midrule
    Current density, $j$ (A m$^{-2}$) & 0-6000 & 1\% \\
    Operating Temperature, $T$ (K) & 893 & 1\% \\
    Anode activation energy, $E_{act,an}$(J mol$^{-1}$) & 53500 & 1\% \\
    Cathode activation energy, $E_{act,cat}$(J mol$^{-1}$) & 77300 & 1\% \\
    Partial pressure, $p_{H_2,an}$ (atm) & 0.6   & 5\% \\
    Partial pressure, $p_{CO_2,an}$ (atm) & 0.15  & 5\% \\
    Partial pressure, $p_{H_2O,an}$ (atm) & 0.25  & 5\% \\
    Partial pressure, $p_{O_2,cat}$ (atm) & 0.08  & 5\% \\
    Partial pressure, $p_{CO_2,cat}$ (atm) & 0.08  & 5\% \\
    Faraday constant, $F$ (C mol$^{-1}$) & 96485 & - \\
    Number of electrons, $n_e$ & 2 & - \\
    Universal gas constant, $R_{gas}$ (J mol$^{-1}$K$^{-1}$) & 8.314 & - \\
    \bottomrule
    \end{tabular}%
  \label{tab:fc_data}%
\end{table*}%

\subsection{General Uncertainty Analysis}

All input parameters in this and following subsections are assumed to follow a univariate normal distribution (no correlation is considered between the parameters). The mean of the normal distribution is the nominal value, while the standard deviation is the parameter uncertainty, which are both indicated in Table \ref{tab:fc_data}. No uncertainty is considered in $j$ in this subsection, and the total number of uncertain parameters is 8.  

\begin{figure*}[h!] 
 \centering
  \includegraphics[width=6.0 in]{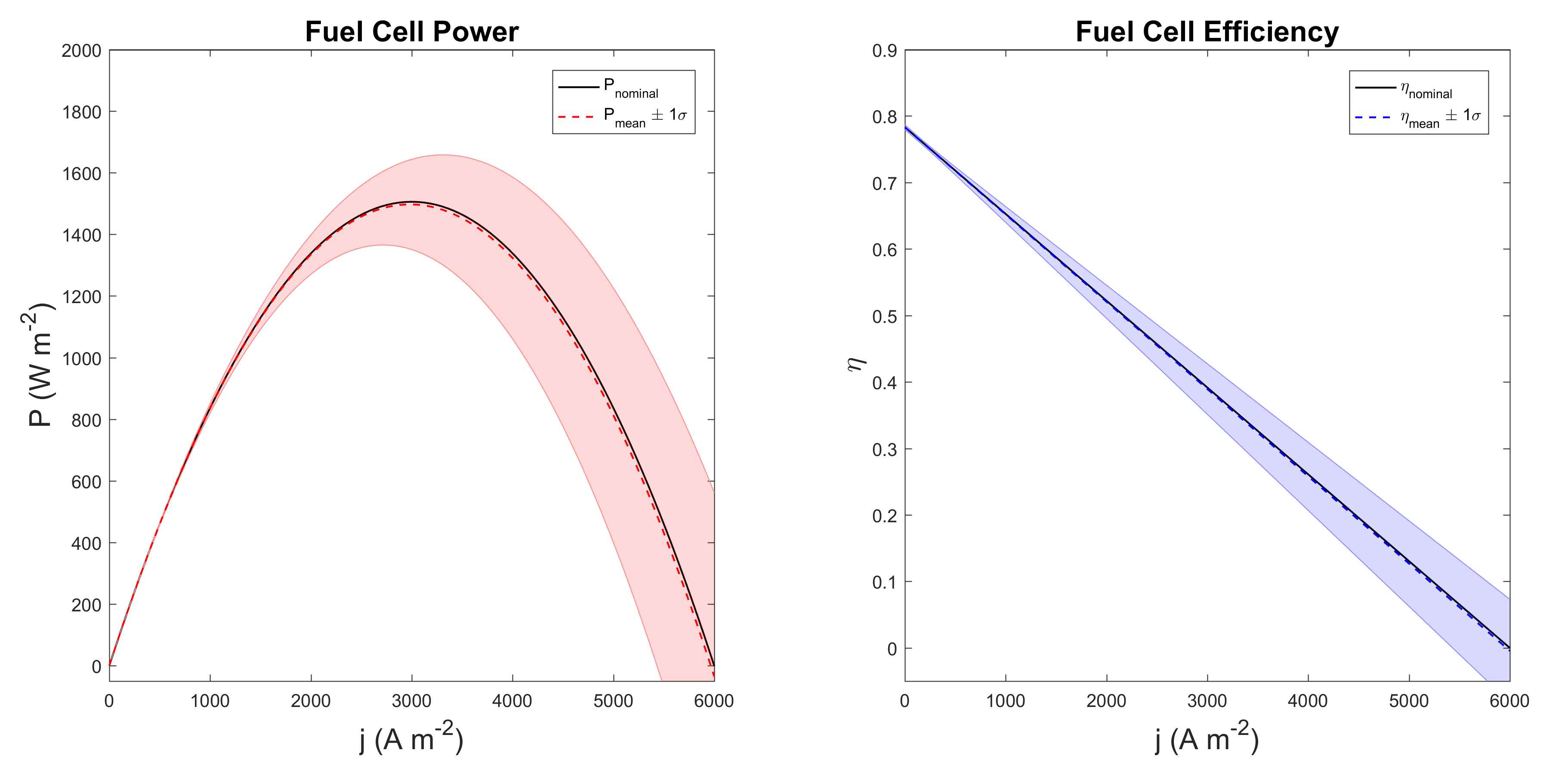}
  \caption{Uncertainty analysis of the MCFC power (left) and efficiency (right) at different current densities}
  \label{fig:FC_gen_UQ}
\end{figure*}

The UQ results of MCFC's power and efficiency are shown in Figure \ref{fig:FC_gen_UQ}. The results that are plotted include: (1) nominal values of $P$ and $\eta$ calculated based on all input parameters held at their nominal values and not perturbed (see Table \ref{tab:fc_data}), (2) the mean value of $P$ and $\eta$ samples after performing Monte Carlo uncertainty propagation, and (3) the 1$\sigma$ uncertainty in $P$ and $\eta$. The results show that the nominal power output starts to increase with $j$ until it reaches a maximum point, after which the power output decreases. The efficiency of the system decreases with increasing $j$. The mean and nominal values of $P$ and $\eta$ are close for all current densities. The results show that the uncertainty in both $P$ and $\eta$ tends to increase as the operating current density increases. It can be observed that the performance of MCFC at higher current densities (4000-6000 A m$^{-2}$) is unreliable, as the relative uncertainty can reach as large as 100\% or even more (i.e. due to the lower power output of these high current regions). In addition, we can observe that operating the MCFC at its maximum power density cannot be done without a certain degree of variability in the power output, which was also observed by \cite{mawardi2006effects} for proton-exchange membrane fuel cells. The maximum power output from the MCFC is $P^* \sim$ 1500 W m$^{-2}$, and it occurs at $j=j^* \sim 3000$ A m$^{-2}$. The system efficiency at $j^*$ is $\eta^* \sim$ 39\% \cite{wu2016performance,wu2017performance}. The relative uncertainty in both $P^*$ and $\eta^*$ is about 10\% of the mean value.

\subsection{Analysis at the Optimum Current Density $j^*$}

In this subsection, the optimum current density $j^*$ is used in the analysis with relative uncertainty of 1\%, where a detailed analysis of $P^*$ (maximum power) and $\eta^*$ (the corresponding efficiency of $P^*$) is presented. The nominal value of the current density is $j=j^*=3000$ W m$^{-2}$. The total number of uncertain parameters explored in the system is 9. 

\begin{figure*}[h!] 
 \centering
  \includegraphics[width=5.5 in]{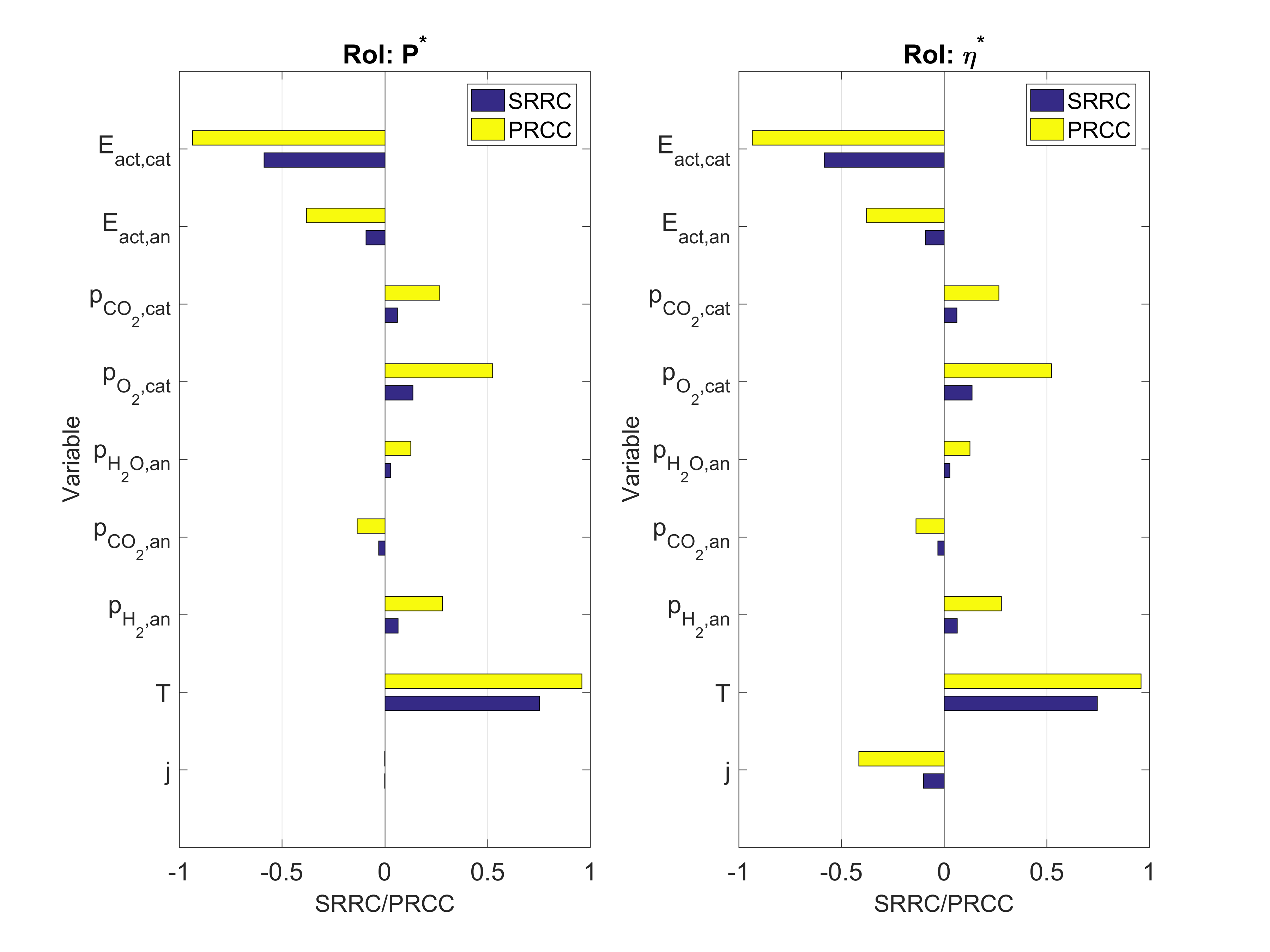}
  \caption{SRRC and PRCC results of the MCFC maximum power (left) and efficiency (right)}
  \label{fig:FC_srrc_prcc}
\end{figure*}

SRRC and PRCC results are presented for power and efficiency output in Figure \ref{fig:FC_srrc_prcc}. The results are generated using the following steps: (1) $10^4$ global random samples are generated from univariate normal distribution (without correlation) for the 9 uncertain parameters indicated in Table \ref{tab:fc_data}, (2) the random samples are all evaluated through the MCFC model described in section \ref{sec:mcfc}, (3) a linear model is constructed and analysis on the rank of the input parameters is performed based on the input-output pairs to calculate SRRC/PRCC.  The results show almost identical behaviour for the $P^*$ and $\eta^*$'s sensitivity, except for $j$.  At the optimum point, the current density shows more negative sensitivity on $\eta^*$ (see Figure \ref{fig:FC_gen_UQ} on the right) compared to $P^*$. This can be justified by the small sensitivity of power to $j$ at the optimum point. According to Figure \ref{fig:FC_gen_UQ} (left), the power output seems to be invariant in the $j$ range of 2800-3200 W m$^{-2}$ which is equivalent to $\pm 6\%$ around the nominal value (i.e. 3000 W m$^{-2}$). This invariance range of 6\% is larger than the uncertainty of 1\% expected for the current density. Therefore, larger sensitivity of $j$ is expected when analyzing the system at points farther from the peak/optimum point (see Figure \ref{fig:FC_gen_UQ} on the left). The results of SRRC/PRCC for both $P^*$ and $\eta^*$ show that the operating temperature $T$ and cathode activation energy $E_{act,cat}$ have the strongest positive and negative sensitivities, respectively, when considering 1\% uncertainty in these two parameters.  

\begin{figure*}[h!] 
 \centering
  \includegraphics[width=5.3 in]{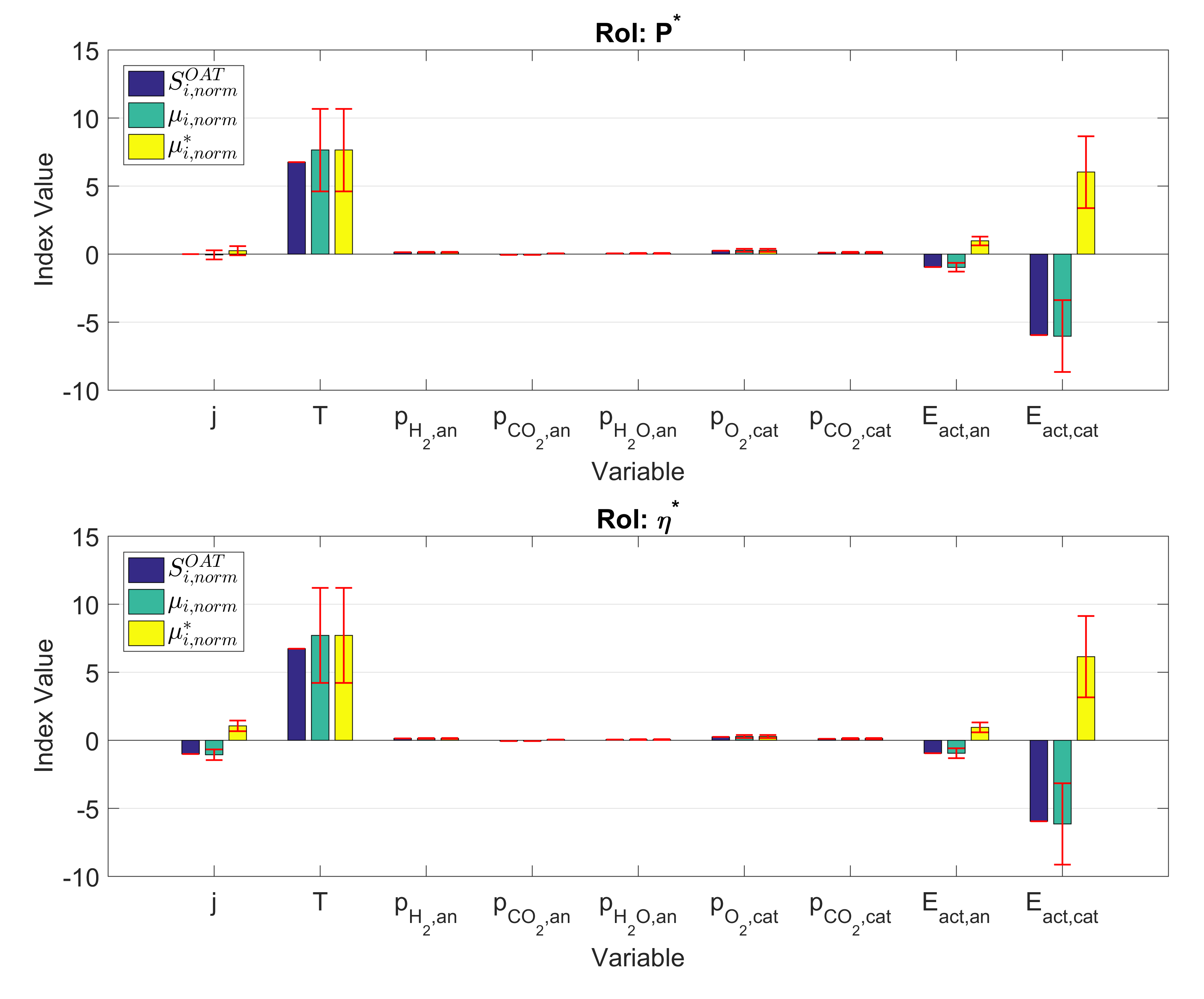}
  \caption{OAT and Morris screening results of MCFC maximum power (top) and efficiency (bottom)}
  \label{fig:FC_morris}
\end{figure*}

Results of OAT and Morris screening for MCFC are plotted in Figure \ref{fig:FC_morris}. OAT results are generated based on a +1\% perturbation around the nominal value, while Morris results are generated using 100 trajectories with 20 grid spaces for each input parameter. The lower/upper bounds for Morris are taken to be  $\pm 3\sigma$ of each parameter (i.e. the physical range of normal distribution). Good agreement between OAT and Morris can be observed for almost all parameters with slight differences due to some nonlinearity/interactions in the model. The following observations can be found from Figure \ref{fig:FC_morris}:
\begin{itemize}
    \item Both $\mu$ and $\mu^*$ have identical estimates (in absolute value) for all parameters except $j$ due to the non-monotonicity of the current density around the peak. 
    \item OAT and Morris results agree in ranking the operating temperature $T$ and cathode activation energy $E_{act,cat}$ as most sensitive parameters. 
    \item A disagreement in ranking the third sensitive parameter can be observed (e.g. for $P^*$), as OAT and Morris identify $E_{act,an}$ as the third sensitive parameter, while SRRC/PRCC gives higher coefficient to $p_{O_2,cat}$. This disagreement is resulted from the uncertainty assigned to each parameter which is considered by SRRC/PRCC calculations only. 
    \item Large uncertainty in the Morris elementary effects for $T$ and $E_{act,cat}$ can be observed due to the nonlinearity and interactions associated with these parameters.
    \item The results in Figure \ref{fig:FC_morris} represent parametric sensitivity only and parametric uncertainty is not yet accounted for. The addition of parametric uncertainty is expected to change the importance ranking of the parameters as will be shown later in this section. 
\end{itemize}

To combine OAT and Morris sensitivities with parametric uncertainty, a comparison of the UQ results of $P^*$ and $\eta^*$ is given in Table \ref{tab:FC_unc}. The results include the absolute uncertainty as calculated by: (1) Monte Carlo uncertainty propagation of the input parameters' uncertainty, (2) deterministic-based UQ using sensitivity coefficients obtained from finite difference OAT (i.e. $S_i^{OAT}$), and (3) deterministic-based UQ using sensitivity coefficients obtained from Morris screening (using $\mu^*$). In general, there is good agreement between the results of the three methods, but with better agreement between Morris and Monte Carlo results. This implies that Morris elementary effects are more accurate than OAT sensitivity. In general, we can observe that Monte Carlo results are bounded by OAT and Morris screening results. OAT tends to underestimate the uncertainty as their sensitivity coefficients are limited and can miss some uncertainty information. On the other hand, Morris' $\mu^*$ seems to overestimate the parameter sensitivity. The cost of Monte Carlo UQ is $10^4$ samples/model evaluation, Morris screening UQ is $10^3$ evaluations, while for OAT is only 10 evaluations. These differences in computational cost are less important here since the model is relatively cheap to evaluate, but these cost differences would be important when applying these methods for computationally-intensive models.

\begin{table}[htbp]
  \centering
  \caption{Absolute uncertainty calculated in the MCFC's maximum power output ($P^*$) and efficiency ($\eta^*$) using different methods}
    \begin{tabular}{ccc}
    \toprule
    Method & $\sigma_{P^*}$ (W m$^{-2}$) & $\sigma_{\eta^*}$ \\
    \midrule
    Monte Carlo UQ    & 144.8 & 0.038 \\
    Deterministic UQ (OAT)  & 138.4 & 0.036 \\
    Deterministic UQ (Morris) & 147.1 & 0.038 \\
    \bottomrule
    \end{tabular}%
  \label{tab:FC_unc}%
\end{table}%

\begin{figure*}[h!] 
 \centering
  \includegraphics[width=5.3 in]{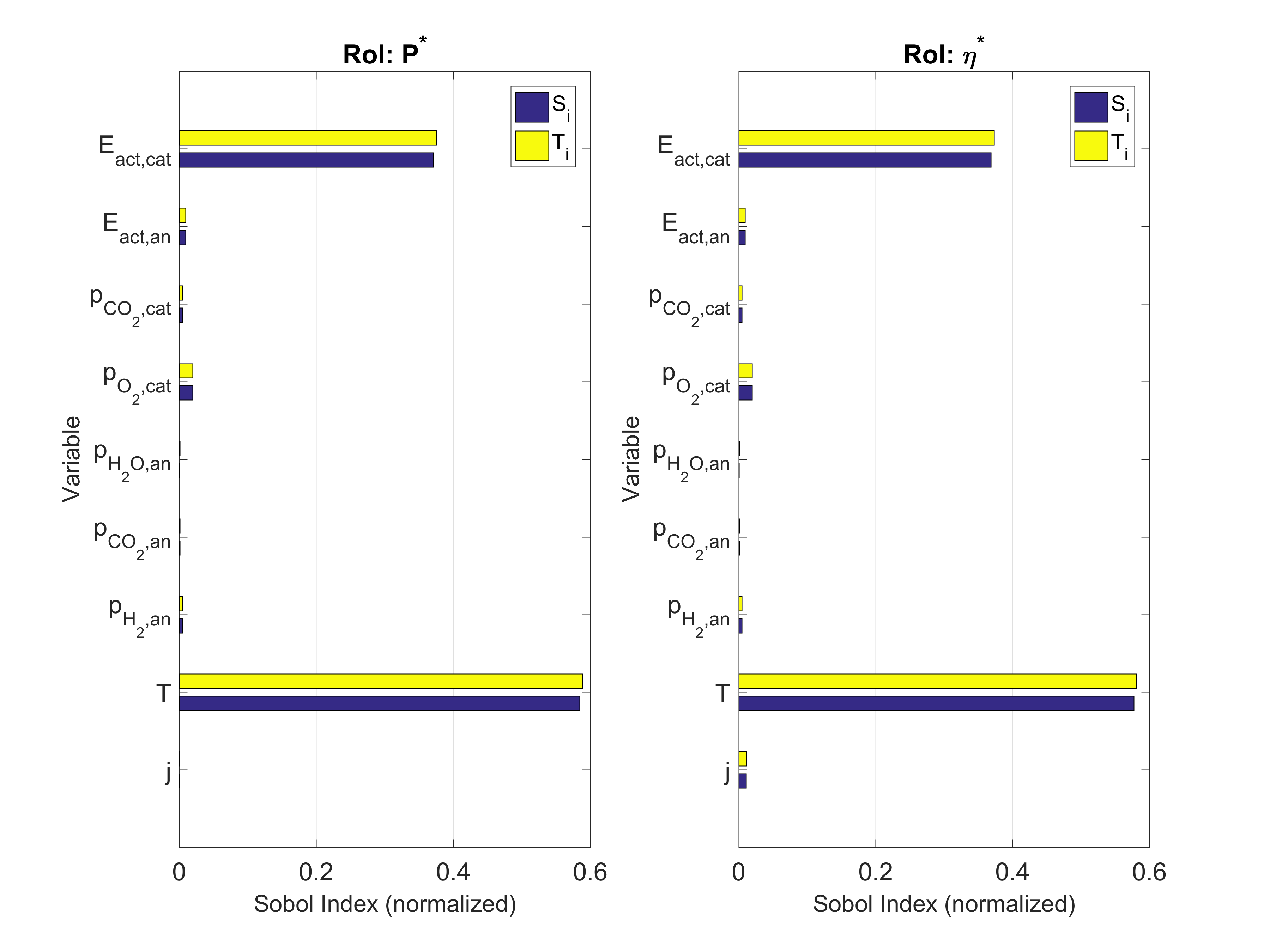}
  \caption{First order and total Sobol indices for MCFC maximum power (left) and efficiency (right)}
  \label{fig:FC_sobol}
\end{figure*}

Application of Sobol indices to decompose the total uncertainty in Table \ref{tab:FC_unc} into the corresponding parameters is given in Figure \ref{fig:FC_sobol}. The Sobol indices reported are normalized, which means that they are expressed as a fraction of the total variance. A first look at the first and total indices shows that the interaction between the parameters has an insignificant effect on the output uncertainty (i.e. $S_i \simeq T_i$) for all parameters. Sobol decomposition agrees with the importance ranking of SRRC and PRRC, as both $T$ and $E_{act,cat}$ have much larger contribution than partial pressure variables. However, Sobol indices highlight the dominance of these two parameters on the power and efficiency variances at the optimum point more than any other method. Both $T$ and $E_{act,cat}$ contribute to more than 90\% of the total variance in $P^*$ and $\eta^*$, implying the high importance of accurately measuring these two parameters if MCFC is operating at its optimum current. The partial pressure $p_{O_2,cat}$ with the highest sensitivity contributes only to about 2\% of the total variance. Therefore, the 1\% uncertainty in $T$ causes a much larger effect on the output variance than the 5\% uncertainty in $p_{O_2,cat}$. It is important noticing that these conclusions are subjective to the reported uncertainties used in Table \ref{tab:fc_data}, meaning that using different uncertainties could change both the variance contribution and the importance ranking.  

A comparison of all methods (OAT, Morris, SRRC, PRCC, and Sobol)  using their importance ranking for the 9 input parameters of the MCFC is given in Table \ref{tab:FC_rank}. All reported results are in absolute value. These methods can be decomposed into two main categories based on their ranking: (1) sensitivity-only methods: OAT and Morris which closely agree on the ranking, and (2) sensitivity and uncertainty methods: SRRC, PRCC, and Sobol indices, which also rank the 9 parameters similarly. In general, all methods agree on ranking the first two sensitive parameters as well as in ranking most of the partial pressure variables. Obvious disagreement between sensitivity-only methods and the uncertainty methods can be found in ranking $E_{act,an}$ and $p_{O_2,cat}$, at which uncertainty methods ranked $p_{O_2,cat}$ as number 3. This came from the difference in the uncertainty assigned to $p_{O_2,cat}$ (5\%), which is higher than that for $E_{act,an}$ (1\%). Therefore, this difference highlights the importance of using these methods together for importance ranking. The two deterministic methods highlight the parametric sensitivity only on the system without considering the parametric uncertainty. For Morris, we reported the value of $\mu_{i,norm}$ instead of $\mu_{i,norm}^*$ to avoid misleading $j$ importance. Using $\mu_{i,norm}^*$ for $j$ will place it on number 5 even though we know that this high $\mu_{i,norm}^*$ originates from the non-monotonic behaviour of $j$ at the peak power, not because $j$ is actually important. However, using $\mu_{i,norm}$ results in a total agreement between the five methods on $j$ ranking. These conclusions imply the importance of using these methods in tandem to obtain a comprehensive understanding of the system being analyzed. Therefore, based on these SA and UQ results, we can conclude the importance of the operating temperature and cathode activation energy on the overall system performance of the MCFC, based on the optimum current point and the prescribed uncertainties used.  

\begin{table*}[htbp]
  \begin{threeparttable}
  \centering
  \small
  \caption{Summary of importance ranking (in absolute value) of the MCFC parameters using different methods used in this study for maximum power output}
    \begin{tabular}{llll|llllll}
    \toprule
    \multicolumn{4}{c|}{Sensitivity-only (Normalized)} & \multicolumn{6}{c}{Sensitivity + Uncertainty} \\
    \midrule
    \multicolumn{2}{c}{OAT ($S_{i,norm}^{OAT}$)} & \multicolumn{2}{c|}{Morris ($\mu_{i,norm}^*$)} & \multicolumn{2}{c}{SRRC} & \multicolumn{2}{c}{PRCC} & \multicolumn{2}{c}{Sobol ($T_i$)} \\
    \midrule
    $T$   & 6.76E+00 & $T$   & 7.66E+00 & $T$   & 7.53E-01 & $T$   & 9.60E-01 & $T$   & 5.89E-01 \\
    $E_{act,cat}$ & 5.94E+00 & $E_{act,cat}$ & 6.06E+00 & $E_{act,cat}$ & 5.88E-01 & $E_{act,cat}$ & 9.36E-01 & $E_{act,cat}$ & 3.75E-01 \\
    $E_{act,an}$ & 9.36E-01 & $E_{act,an}$ & 9.50E-01 & $p_{O_2,cat}$ & 1.36E-01 & $p_{O_2,cat}$ & 5.24E-01 & $p_{O_2,cat}$ & 1.98E-02 \\
    $p_{O_2,cat}$ & 2.64E-01 & $p_{O_2,cat}$ & 2.81E-01 & $E_{act,an}$ & 9.13E-02 & $E_{act,an}$ & 3.82E-01 & $E_{act,an}$ & 9.28E-03 \\
    $p_{H_2,an}$ & 1.29E-01 & $p_{H_2,an}$ & 1.34E-01 & $p_{H_2,an}$ & 6.45E-02 & $p_{H_2,an}$ & 2.80E-01 & $p_{H_2,an}$ & 4.72E-03 \\
    $p_{CO_2,cat}$ & 1.25E-01 & $p_{CO_2,cat}$ & 1.30E-01 & $p_{CO_2,cat}$ & 6.14E-02 & $p_{CO_2,cat}$ & 2.68E-01 & $p_{CO_2,cat}$ & 4.54E-03 \\
    $p_{CO_2,an}$ & 5.52E-02 & $p_{CO_2,an}$ & 5.62E-02 & $p_{CO_2,an}$ & 2.99E-02 & $p_{CO_2,an}$ & 1.34E-01 & $p_{CO_2,an}$ & 9.16E-04 \\
    $p_{H_2O,an}$ & 4.77E-02 & $p_{H_2O,an}$ & 5.32E-02 & $p_{H_2O,an}$ & 2.83E-02 & $p_{H_2O,an}$ & 1.27E-01 & $p_{H_2O,an}$ & 8.45E-04 \\
    $j$   & 1.03E-02 & $j$   & 5.00E-02$^\ddagger$ & $j$   & 3.58E-05 & $j$   & 1.62E-04 & $j$   & 5.16E-04 \\
    \bottomrule
    \end{tabular}%
    \begin{tablenotes}
      \small
      \item $\ddagger$ The value of $|\mu_{i,norm}|$ is used for $j$ instead of $\mu_{i,norm}^* =$ 1.24E-01.
    \end{tablenotes}
  \label{tab:FC_rank}%
  \end{threeparttable}
\end{table*}%

Finally, it is worth mentioning that all these SA and UQ results are generated in an automated way based on a framework developed by the authors in R language, which is also expected to be available in Python as well. Currently, the package accepts the energy model in an analytical form, uncertainties of the input parameters, and other important sampling and user inputs. The package performs all SA and UQ methods and provide final results to the user. Our next efforts will focus on extending these methods to accept more advanced energy models that come from computer simulations and transient analysis.   

\section{Conclusions}
\label{sec:summary}

In this study, various methods of sensitivity and uncertainty analyses have been introduced and applied in the context of energy systems. This paper has described the theory and methods behind performing detailed SA and UQ of mathematical and engineering models in a comprehensive form. The implementation in this paper facilitates SA and UQ processes and integrate them within the model. SA methods include local methods using finite difference (OAT perturbation) which provides a quick but shallow understanding of the system sensitivity. Morris screening is an example of a screening method which can give an indication of parameter sensitivity as well as its nonlinearity or interaction behaviour with other parameters. Standardized regression and partial correlation coefficients are used to perform regression-based SA, which measure the strength of linear correlation between the input and the output. Two methods for UQ are introduced, the first is Monte Carlo-based uncertainty propagation, while the other is deterministic-based using sensitivity profiles provided from OAT or Morris screening methods. Finally, Sobol indices are utilized to decompose the total response variance into portions attributable to each input parameter. Sobol indices are preferred to assess models that are complex, nonlinear, non-monotonic, and have excessive parameter interactions. As a major objetive of this paper is to give the reader a detailed understanding of the methods, both simple and benchmark functions (e.g. Ishigami, Morris, Sobol) in a form of numerical tests are used first to compare the methods. A practical test is also performed on a standalone fuel cell model (MCFC) which has been rarely investigated in previous studies. The results from the MCFC analysis show that the uncertainty in both the optimum power output and its corresponding system efficiency are about 10\%. SA and UQ final results demonstrate that the operating temperature and cathode activation energy are the most influential parameters for the MCFC, as they are responsible of more than 90\% of the total power and efficiency variance. In future work, applications to more advanced hybrid energy systems featuring fuel cells will be conducted. These systems could be for example a hybrid system of fuel cell, thermoelectric generator, and regenerator. Parametric sensitivity and uncertainty will be analyzed for such systems and conclusions about their performance will be drawn. On the development side, extension of the capabilities of SA and UQ to account for simulation models that do not have a closed mathematical form will be performed. In these cases, some reduced order modeling and machine learning techniques (e.g. regression, Gaussian processes, neural networks, etc.) can be used if the model is expensive to evaluate.  

\phantomsection
\bibliographystyle{elsarticle-num}
\small
\bibliography{references}



\end{document}